\documentclass[prl,twocolumn,showpacs,superscriptaddress,amssymb]{revtex4}

\usepackage{psfrag,graphicx}
\usepackage{dcolumn}
\usepackage{bm}
\usepackage{amsfonts,amssymb,amsmath}        

\usepackage{xcolor}

\newcommand{\be}{\begin{equation}}
\newcommand{\ee}{\end{equation}}
\newcommand{\bq}{\begin{eqnarray}}
\newcommand{\eq}{\end{eqnarray}}

\newcommand{\ii}{\boldsymbol{i}}

\newcommand{\pp}{\boldsymbol{p}}

\definecolor{kon}{rgb}{.8,.1,0.3}

\begin{document}

\title{Two-dimensional Chern semimetals on the Lieb lattice} 

\author{Giandomenico Palumbo}
 \affiliation{School of Physics and Astronomy, University of Leeds, Leeds, LS2 9JT, United Kingdom}
\author{Konstantinos Meichanetzidis}
 \affiliation{School of Physics and Astronomy, University of Leeds, Leeds, LS2 9JT, United Kingdom}

\date{\today}

\pacs{71.10.Fd, 03.65.Vf, 71.10.Pm, 71.20.Gj}

\begin{abstract}
In this work, we propose a new and simple model that supports Chern semimetals. These new gapless topological phases share several properties with the Chern insulators like a well-defined Chern number associated to each band, topologically protected edge states and topological phase transitions that occur when the bands touch each, with linear dispersion around the contact points. The tight-binding model, defined on the Lieb lattice with intra-unit-cell and suitable nearest-neighbor hopping terms between three different species of spinless fermions, supports a single Dirac-like point.  The dispersion relation around this point is fully relativistic and the $3\times3$ matrices in the corresponding effective Hamiltonian satisfy the Duffin-Kemmer-Petiau algebra. We show the robustness of the topologically protected edge states by employing the entanglement spectrum. Moreover, we prove that the Chern number of the lowest band is robust with respect to weak disorder. For its simplicity, our model can be naturally implemented in real physical systems like cold atoms in optical lattices.
\end{abstract}

\maketitle

{\bf \em Introduction:--} Topological phases represent one of most exciting and interesting field of condensed matter physics. Topological insulators and superconductors are well-known examples of free fermion systems defined by a gapped bulk that support robust gapless edge states \cite{Ryu}. All these systems described by non-interacting Hamiltonians fit in the periodic table of fermionic topological phases and the topological phase transitions occur when the bulk gap closes. In particular, two-dimensional chiral topological systems like the Haldane model \cite{Haldane} and p-wave superconductors \cite{Read} are characterized by the topological Chern number that fixes the number of topologically protected edge modes. The former is an example of topological insulator in the class A, called also Chern insulator, where both time-reversal and particle-hole symmetries are broken while the latter is a topological superconductor living in the class D, where particle-hole symmetry is instead preserved.\\
In the last years, part of the research of new topological phases has focused on gapless bulk systems like three-dimensional Weyl \cite{Burkov} and Dirac semimetals \cite{Kane} that represent three-dimensional versions of graphene. In these systems, the bands touch each other in a discrete set of points that can be seen as point-like defects in momentum space. These Weyl and Dirac points are topologically protected by well-defined Berry phases and the boundaries support suitable gapless modes.
Clearly, when the time-reversal symmetry is broken in two dimensions, the Chern number can characterize the bands of the semimetals only when those bands do not touch each other, i.e. when there are no Dirac or Weyl cones. Some extended Haldane models with these characteristics have been recently analyzed \cite{Goldman, Spielman}.\\
\begin{figure}[htp]
\begin{center}
\includegraphics[scale=0.22]{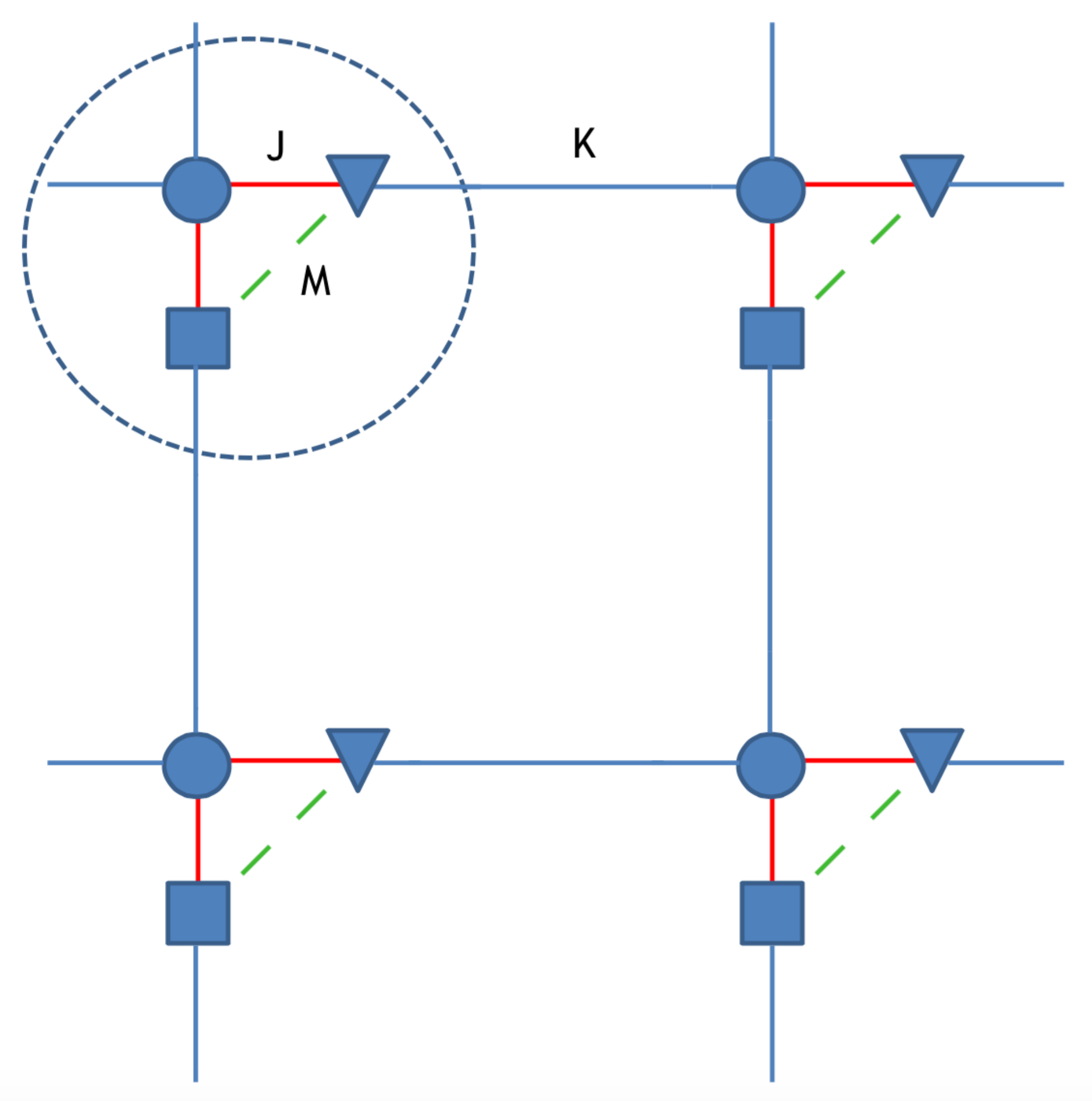}\\
\includegraphics[scale=0.17]{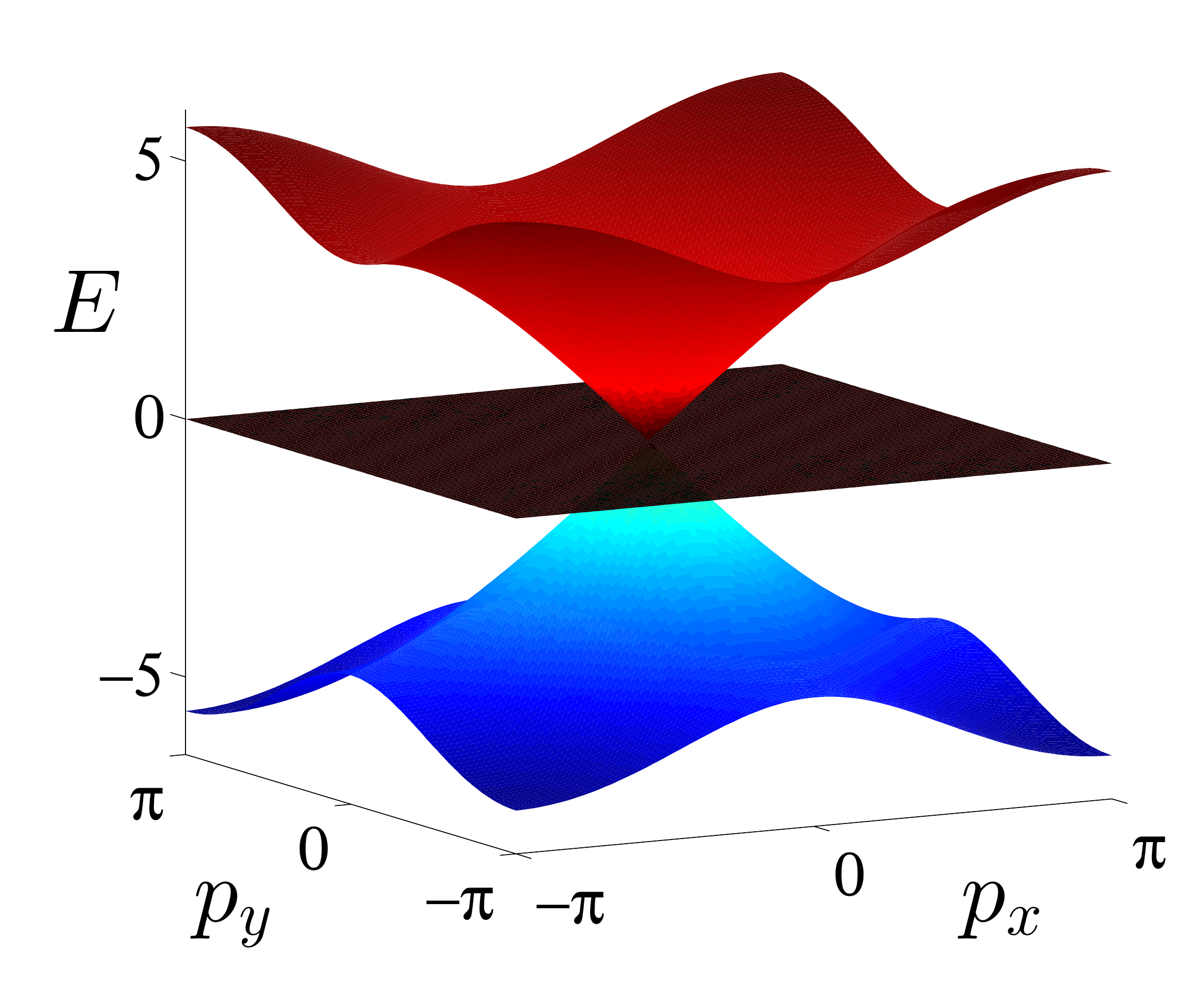}
\includegraphics[scale=0.17]{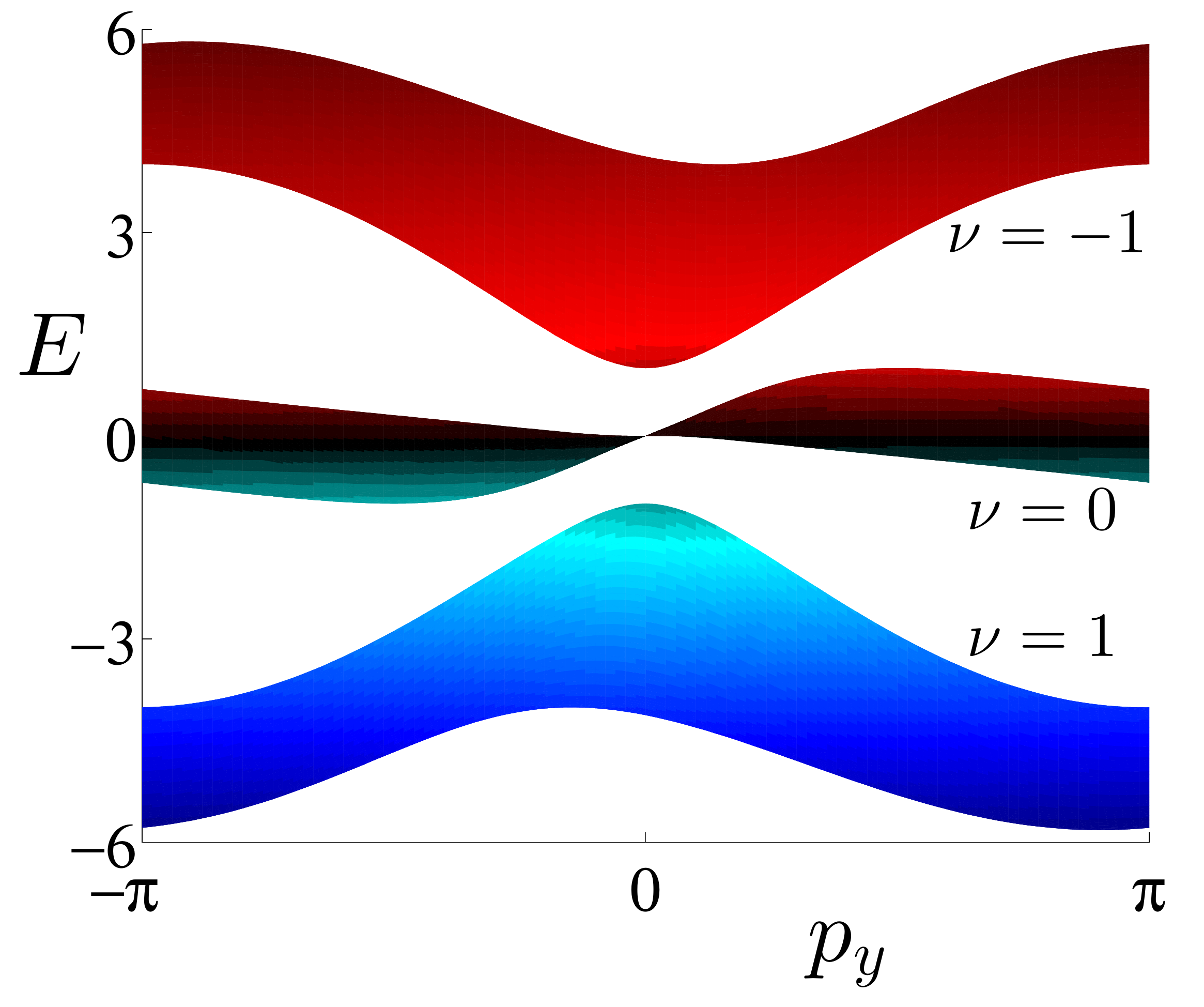}
\end{center}
\caption{\label{Fig1} (Up) This picture shows a square lattice with three species of fermions represented by triangles, circles and squares. The dashed circle encloses the unite cell. Intra-unit-cell hopping terms are represented by red lines, the nearest-neighbor ones with blue lines. The dashed green line identifies the hopping between non-adjacent fermions. (Low-left) Dispersion energy for $K=-J=1$ and $M=0$. (Low-right) Projection of dispersion energy for with $M=0.5 i$.}
\end{figure}

The goal of this work is to present a new and simple model that supports two-dimensional Chern semimetallic phases, which share several properties with the Chern insulators. Our tight-binding model is defined on a Lieb lattice with intra-unit-cell and suitable nearest-neighbor hopping terms between three species of free spinless fermions. The model supports only a Dirac-like cone due to the presence of a zero-energy flat band in the middle. In fact, it is possible to avoid the femion doubling in the lattice if a flat band is present. At this point, it is possible to deform the bands by introducing a further hopping term between non-adjacent fermions in the unit cell.
In this case, as we show in the next section, the lower band is characterized by a non-zero Chern number $\nu=\pm 1$ and the model supports robust edge states. We analyze the edge modes by the employing entanglement spectrum and we demonstrate the robustness of the Chern number with respect to the presence of weak disorder.
Moreover,  we show that the corresponding effective Hamiltonian $h_{\mathrm{eff}}$ is fully relativistic but different with respect to a Dirac Hamiltonian, because the $3\times3$ matrices in $h_{\mathrm{eff}}$ satisfies the Duffin-Kemmer-Petiau algebra \cite{Nieto}. This implies that our model does not fit in any already known periodic table of topological gapless phases based on K-theory and Clifford algebra \cite{Ryu2}. Finally, due to the presence of only intra-unit-cell and suitable nearest-neighbor hopping terms, our model can be easily implemented in real physical systems like cold atoms in optical lattices.\\

{\bf \em Lattice model and Chern number:--} To begin with, we introduce our tight-binding model. We consider three species of spinless fermions on a Lieb lattice as shown in Fig. 1, described by the following Hamiltonian
\begin{align}
H =\sum_{\ii}\Big[J(a^{\dagger}_{\ii}b_{\ii}+b^{\dagger}_{\ii}c_{\ii})+ & K (a^{\dagger}_{\ii}b_{\ii+\hat{x}}+b^{\dagger}_{\ii}c_{\ii+\hat{y}})+M c^{\dagger}_{\ii}a_{\ii}\Big] \nonumber \\ +&\mathrm{h.c.},
 \end{align}
where $i$ is the site index, $\hat{x}=(1,0)$, $\hat{y}=(0,1)$, and $a$, $b$ and $c$ are the three species of fermions represented in Fig. 1 in terms of triangles, circles and squares, respectively.
Here, the tunneling coefficients $J$ and $K$ are taken real, while $M=m\, e^{i\theta}$ is complex. By imposing periodic boundary conditions we introduce the Fourier transformation $a_{k,\ii} = \sum _{\pp} e^{i\pp \cdot \ii} a_{k,\pp}$, where $k=1,2,3$ is the species index, to obtain $H = \sum_{\pp} \psi^\dagger_{\pp} h(\pp) \psi_{\pp}$, where 
$\psi_{\pp} = (a_{\pp}, b_{\pp}, c_{\pp})^{T}$, 
$\pp\in \text{BZ}=[-\pi ,\pi)\times[-\pi,\pi)$ and the kernel $h(\pp)$ is a $3\times 3$ hermitian matrix. It is straightforward to see that the time-reversal symmetry is broken, namely $h(\pp) \neq h^{*}(-\pp)$, but is restored when $M$ is real. In particular, when $M=0$, the system supports a single Dirac-like cone and a zero-energy flat band as shown in Fig. 1.
First of all, the presence of the flat band allows to avoid the fermion doubling problem
as shown in \cite{Fradkin}. In this case, our model behaves similarly to other three-band models define on Lieb and Kagome lattices \cite{Chamon, Franz, Goldman2, Smith, Carpentier}. 
\begin{figure}[htp]
\begin{center}
\includegraphics[scale=0.14]{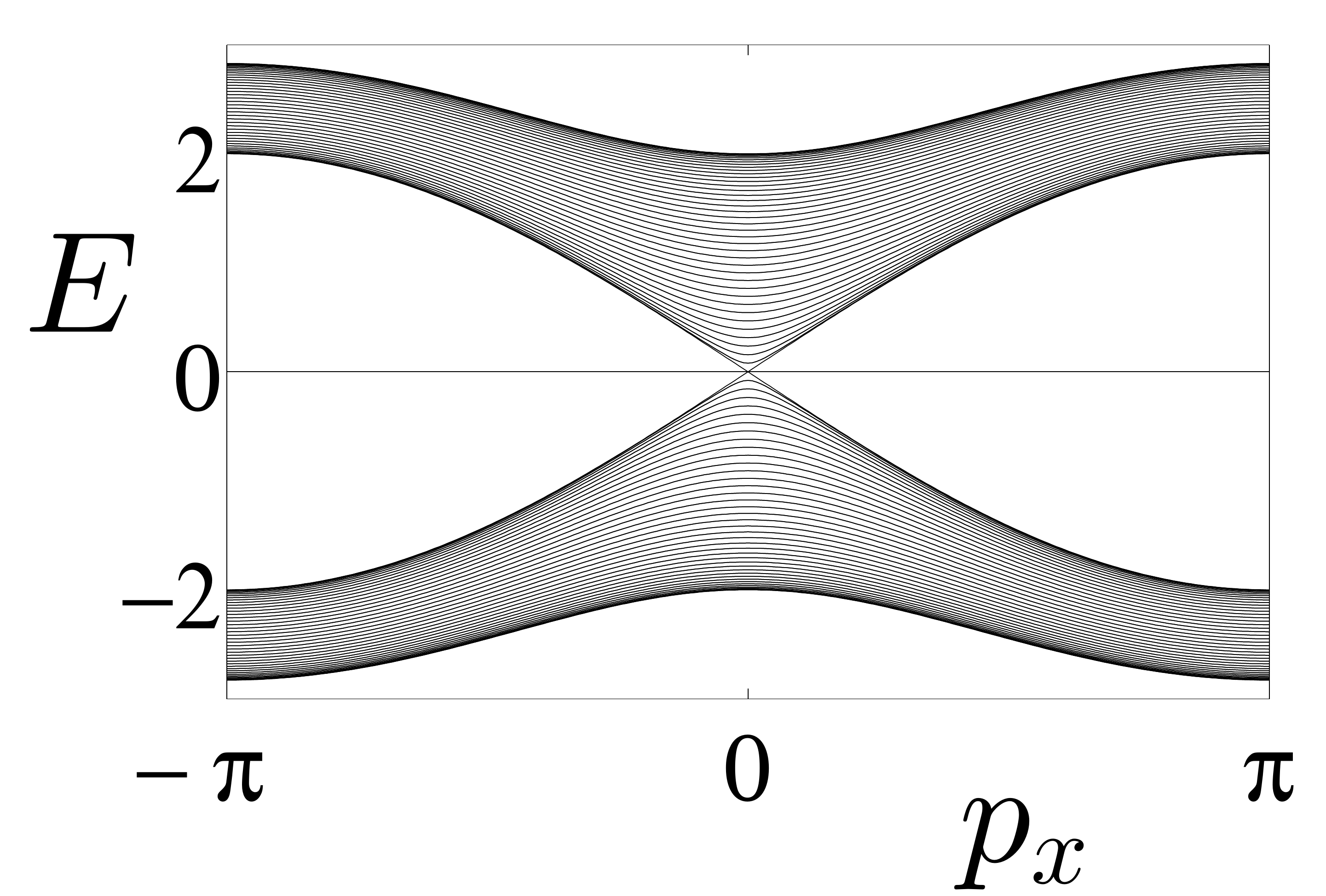}
\includegraphics[scale=0.14]{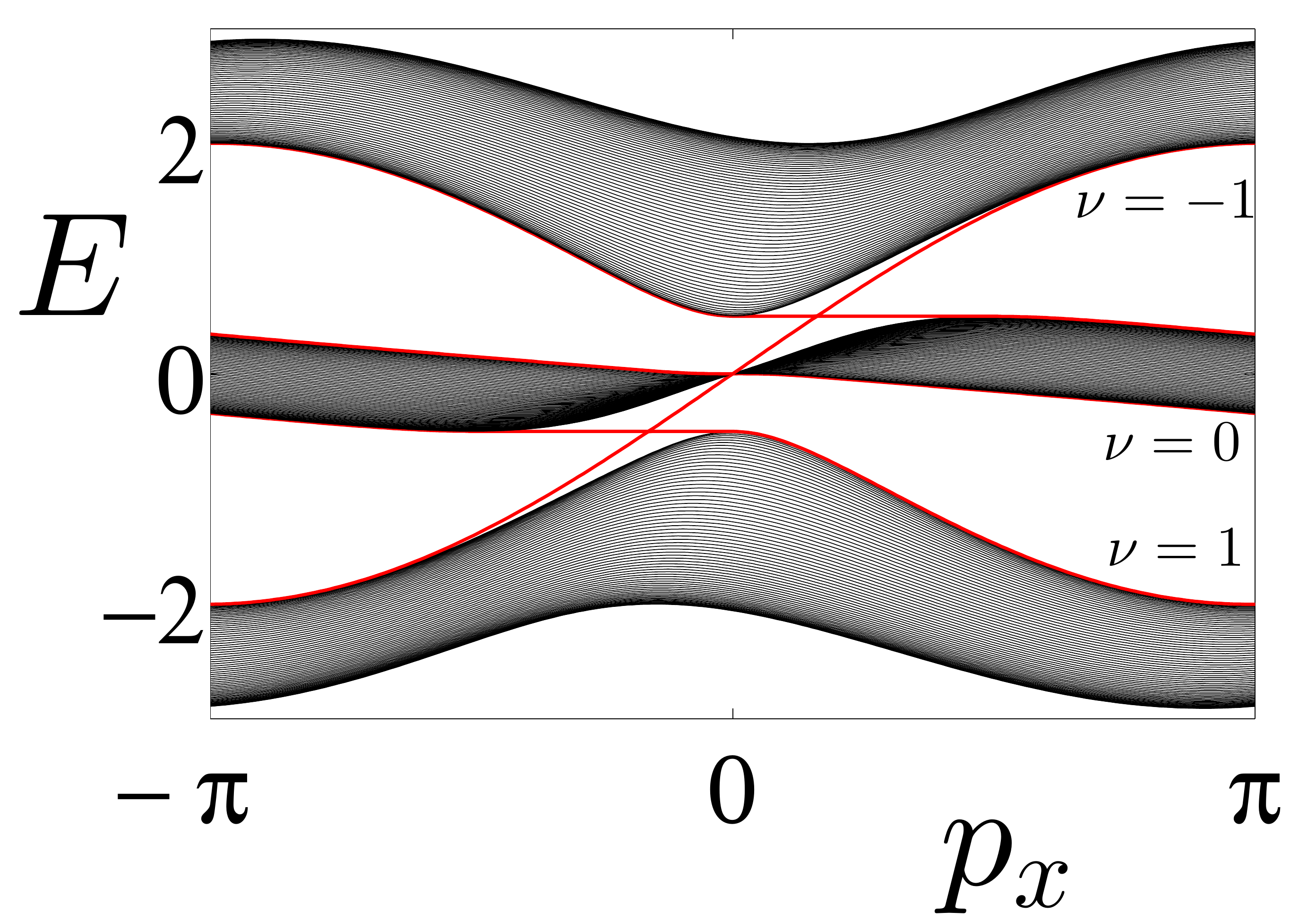}
\includegraphics[scale=0.14]{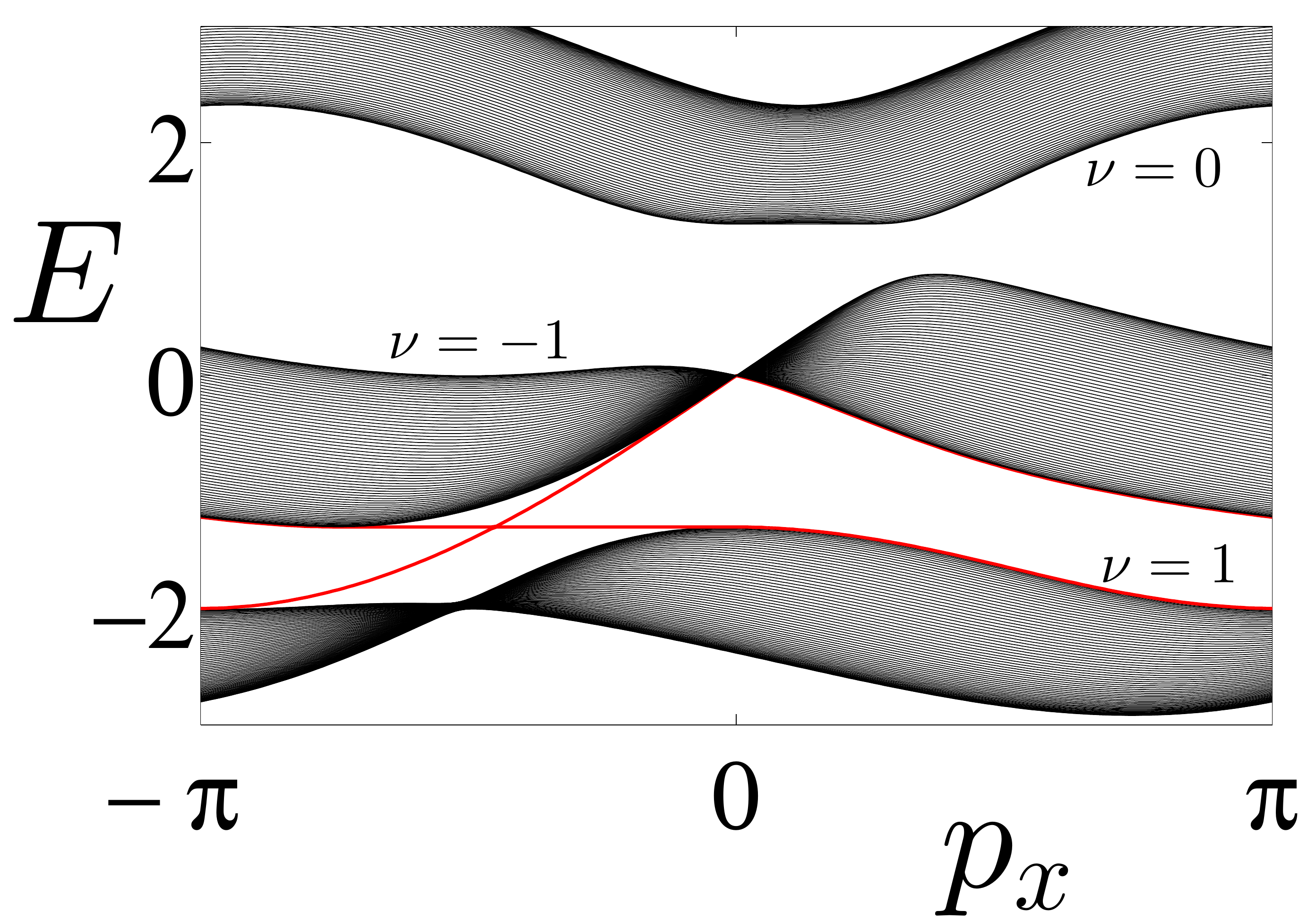}
\includegraphics[scale=0.14]{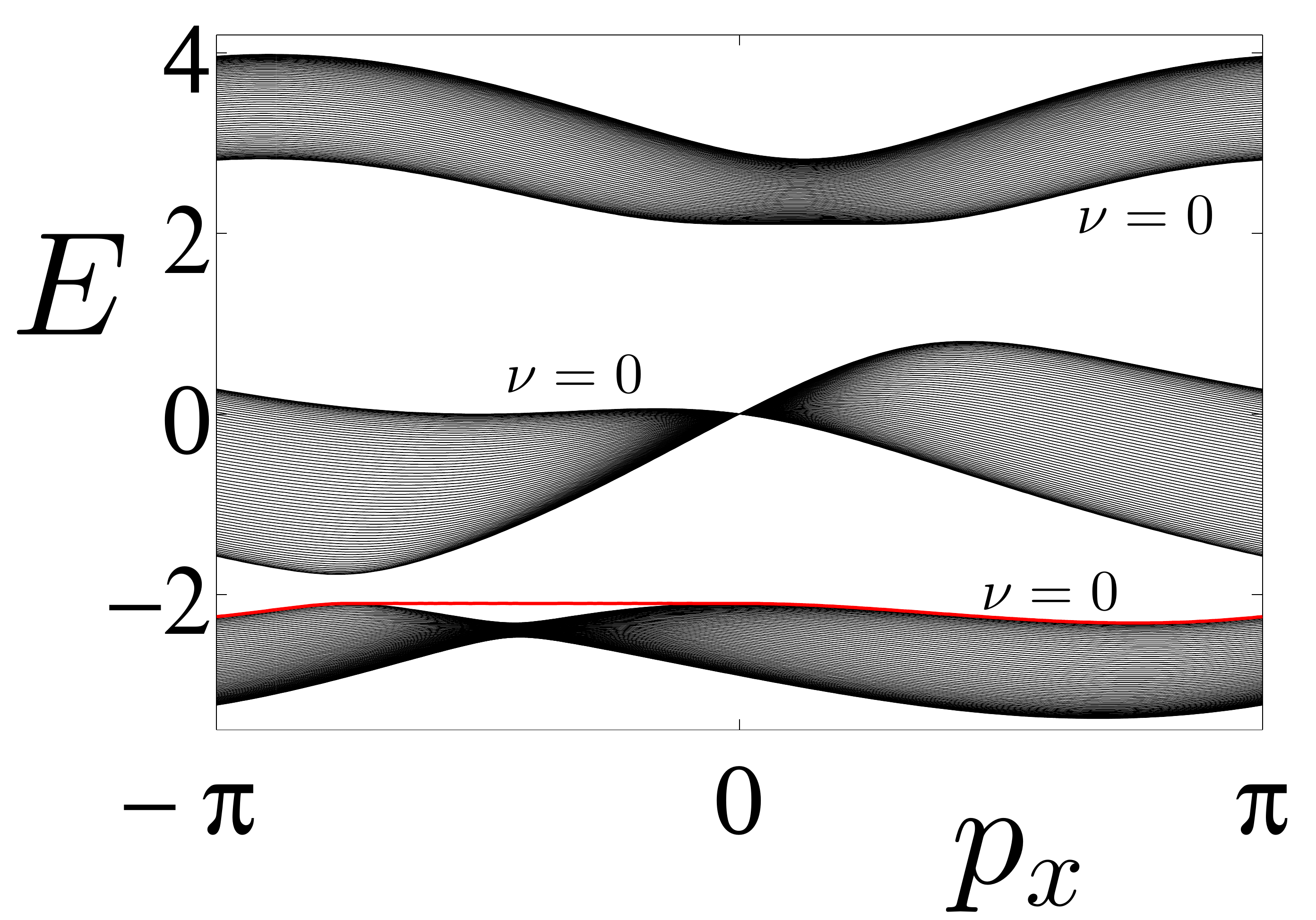}
\end{center}
\caption{\label{Fig2} Energy bands for $K=-J=1$  on cylinder of length $L_y=80$ as a function of momentum $p_x$. (Up-left) $M=0$. A single Dirac-like cone and a zero-flat band appear in the system. (Up-right) $m=0.5$ and $\theta=\pi/2$. Upper and lower bands have non-zero Chern number and the crossing red lines represent the edge modes between them. (Low-left)$m=1.3$ and $\theta=\pi/4$. the upper gap opens and the Chern number is acquired by the middle band after a topological phase transition when the bands touch. (Low-right) When even the lower gap opens, the system becomes a trivial insulator.}
\end{figure}
However, a crucial difference emerges with respect to the latter, when a non-zero complex $M$ is switched on. In this case, the complex hopping term with a small $m<1$ opens a pseudo-gap between the lower and upper bands and deforms the middle bands such that the systems remains gapless as shown in Fig. 1.
We now show that in this semimetallic phase, our model behaves as a non-trivial topological semimetal, where each band has a well-defined Chern number $\nu$. This is indeed possible because the time-reversal symmetry is broken and the bands do not touch each other at any point in the BZ like for example in graphene and Weyl semimetals.
The Chern number related to the n-th band with normalized Bloch wave function $|n(\pp)\rangle$, is defined by
\begin{eqnarray}\label{Chern}
\nu_{n}=\frac{1}{2 \pi\,i}\int_{BZ} d^{2}p\, F_{xy},
 \end{eqnarray}
where the Berry connection $A_{\alpha}$ ($\alpha=x, y$) and the corresponding curvature tensor $F_{xy}$ are given by $A_{\alpha}=\langle n(\pp)| \frac{\partial}{\partial p_{\alpha}}| n(\pp)\rangle$ and $F_{xy}=\partial_{x}A_{y}-\partial_{y}A_{x}$.
Here, we are assuming that there is no degeneracy for the n-th state and we obtain a Chern number $\nu=\pm 1$ by integrating the Berry curvature on the lower band. For our numerical calculations, we use the discretised version of (\ref{Chern}) derived in \cite{Fukui}. From a geometric point of view, the presence of the Chern number is connected to the presence of a non-zero flux passing through the triangles defined by the hoppings $J$ and $M$ inside the unit cell when $M$ is complex. Fig. 3 reproduces the phase diagram of this new topological phase for $K=1$ and $\theta=\pi/2$ and displays how the chirality depends on the sign of $m$ and $J$.
\begin{figure}[htp]
\begin{center}
\includegraphics[scale=0.197]{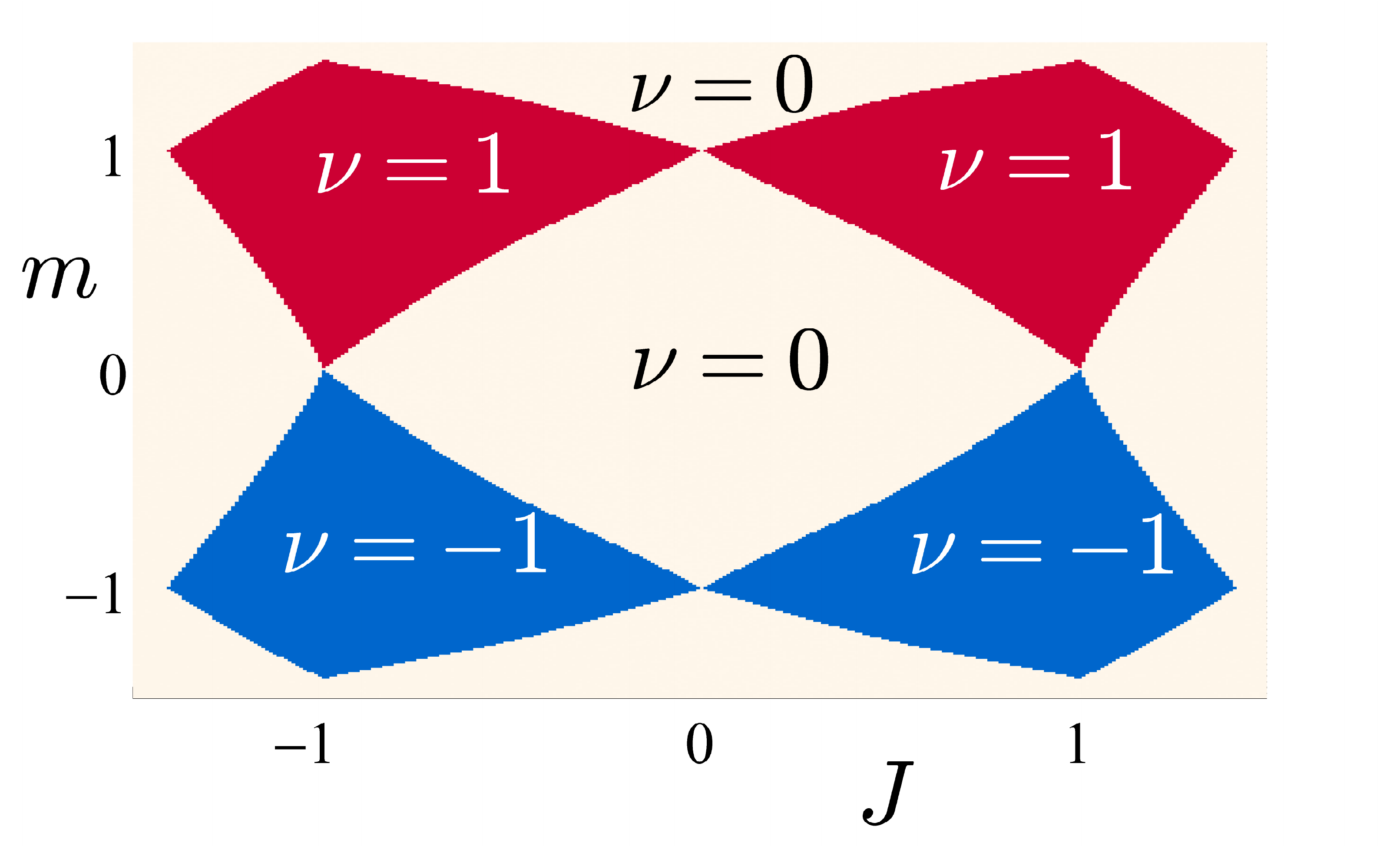}
\includegraphics[scale=0.188]{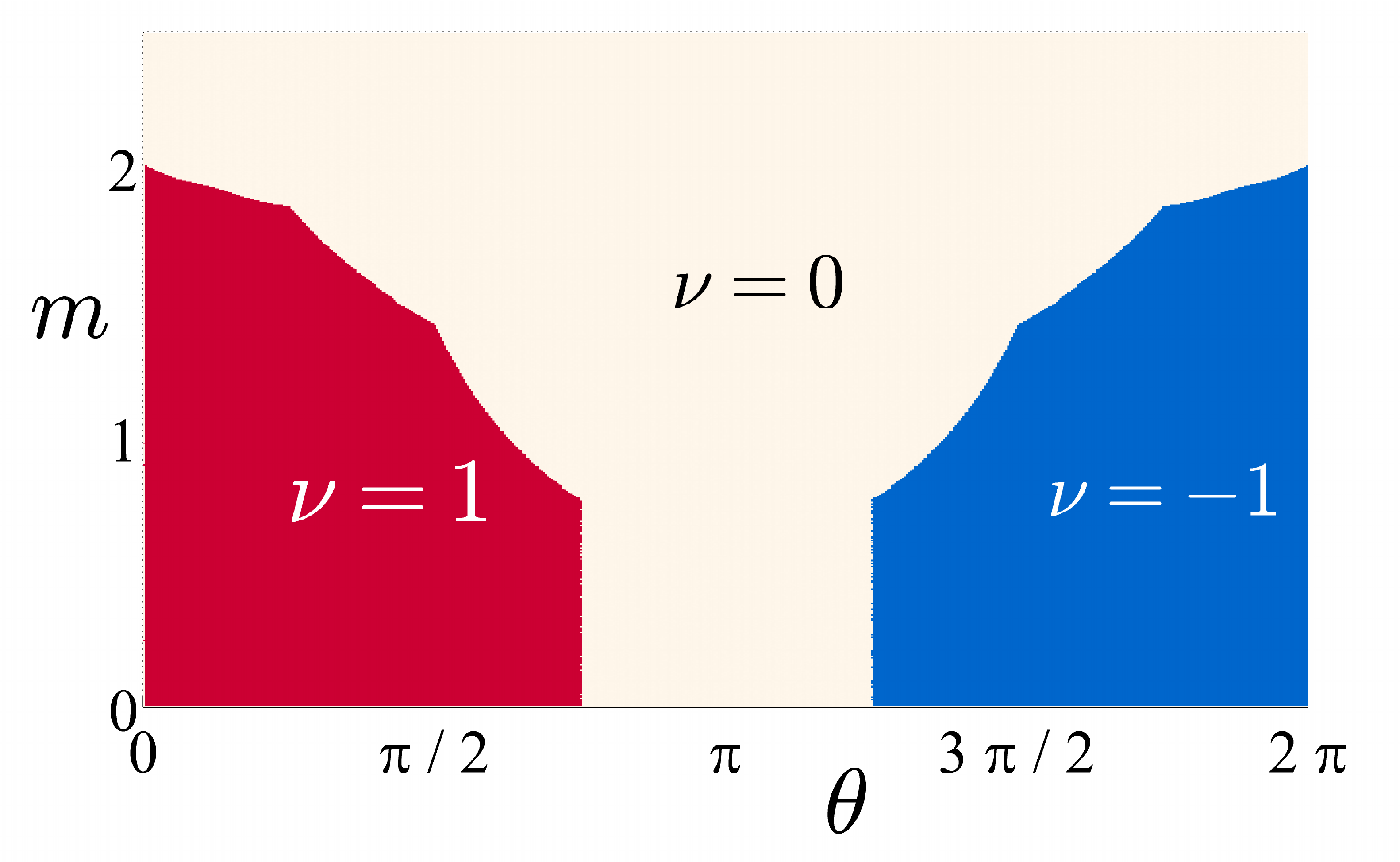}
\end{center}
\caption{\label{Fig3} (Up) Chern number phase diagram in function of $J$ and $m$ with $K=1$ and $\theta=\pi/2$ when the lowest band is completely filled. (Low) Phase diagram in function of $m$ and $\theta$ with $K=-J=1$.}
\end{figure}
Importantly, topological phase transitions occur when the bands touch each other, with linear dispersion around the contact points.
Moreover, as we show in the next section, topologically edge states are associated to the non-zero Chern number. Thus, this topological semimetal shares several properties with the Haldane model but there are also several differences.
Firstly, in the momentum space, there appears only a Dirac-like point. This property is due to the presence of a zero-energy flat band for $M=0$ and is shared with some three-band realizations of Chern insulators, even if in these last cases, the topological phases are obtained only by introducing spin-orbit interactions \cite{Chamon, Franz, Goldman2, Smith}. Secondly, in the low-energy regime, the corresponding effective kernel Hamiltonian $h_{\mathrm{eff}}$ is not defined by any Dirac or Dirac-Weyl Hamiltonian even if there is a single Dirac-like point. This is possible because $h_{\mathrm{eff}}$ is a fully relativistic first-order Hamiltonian, given by
\begin{eqnarray}\label{Duffin-Kemmer}
h_{\mathrm{eff}}=K\left[\beta^{x},\beta^{0}\right]\,p_{x}+K\left[\beta^{y},\beta^{0}\right]\,p_{y}+M\beta^{0},
 \end{eqnarray}
where we have fixed $J=-K$. Here, the $3\times 3$ $\beta^{\mu}$ matrices
\begin{eqnarray}
\beta^{0}=\left(
\begin{array}{ccc}
0 & 0 & -1\\
0 & 0 & 0 \\
1 & 0 & 0 
\end{array}
\right), \hspace{0.8cm}
\beta^{x}=\left(
\begin{array}{ccc}
0 & 0 & 0\\
0 & 0 & i \\
0 & -i & 0 
\end{array}
\right), \nonumber \\ 
\beta^{y}=\left(
\begin{array}{ccc}
0 & -i & 0\\
i & 0 & 0 \\
0 & 0 & 0 
\end{array}
\right),\hspace{2.0cm}
\end{eqnarray}
satisfy the following conditions
\begin{eqnarray}
\beta^{\mu}\beta^{\nu}\beta^{\sigma}+\beta^{\sigma}\beta^{\nu}\beta^{\mu}=
\beta^{\mu}\eta^{\nu\sigma}+\beta^{\sigma}\eta^{\nu\mu}, \nonumber\,\,\,\,\, (\beta^{\mu})^{3}=\eta^{\mu\mu}\beta^{\mu},
 \end{eqnarray}
where $\eta^{\mu\nu}$ is the the relativistic Minkowski metric such that $\mathrm{diag}\,\, \eta^{\mu\nu}=(-1,1,1)$ and there is no summation on repeated indexes. The above relations identify a generalized Clifford algebra, called Duffin-Kemmer-Petiau algebra \cite{Nieto}.
This algebra is associated to the Duffin-Kemmer-Petiau theory that describes relativistic spin-0 and spin-1 particles by employing the same formalism used by Dirac for spin-1/2 particles. In particular, the effective Hamiltonian in (\ref{Duffin-Kemmer}) formally describes spin-0 quasiparticles in two dimensions and the corresponding spinor field satisfies the Klein-Gordon equations. This also implies that our model does not fit in any already known periodic table of topological semimetals because this classification is based on the K-theory associated to the standard Clifford algebra \cite{Ryu2}.\\

{\bf \em Edge states and entanglement spectrum:--} As is well-known, one of the main properties of topological phases concerns the existence of robust edge states. In particular, in two-dimensional time-reversal broken phases, the number of chiral edge modes coincides with the value of Chern number. This is well established when the systems have a gapped bulk but still an open question for Chern semimetals.
In order to show the edge state energy dispersion we impose open boundary conditions in the $y$-direction and periodic in the $x$-direction we perform a Fourier transformation that decomposes our two-dimensional Hamiltonian into decoupled one-dimensional Hamiltonians describing chains of length $L_y$ parametrised by $p_x$.
In Fig. 2 it is clear that edge states appear and cross between bands with non-trivial Chern numbers of opposite sign.
Here, we show that Chern semimetals have topologically protected edge modes by employing the entanglement spectrum \cite{Haldane2}.
\begin{figure}[htp]
\begin{center}
\includegraphics[scale=0.15]{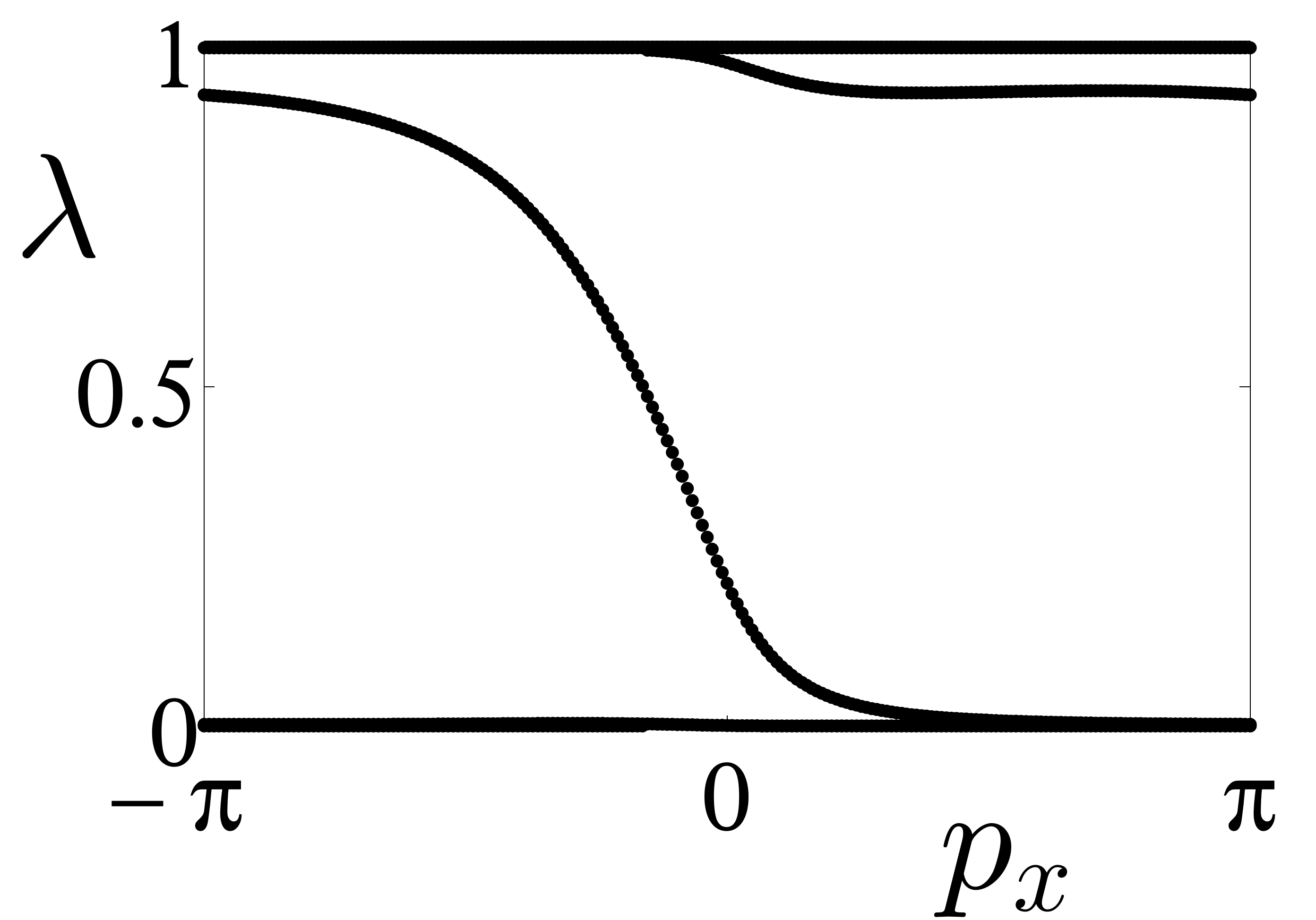}
\includegraphics[scale=0.15]{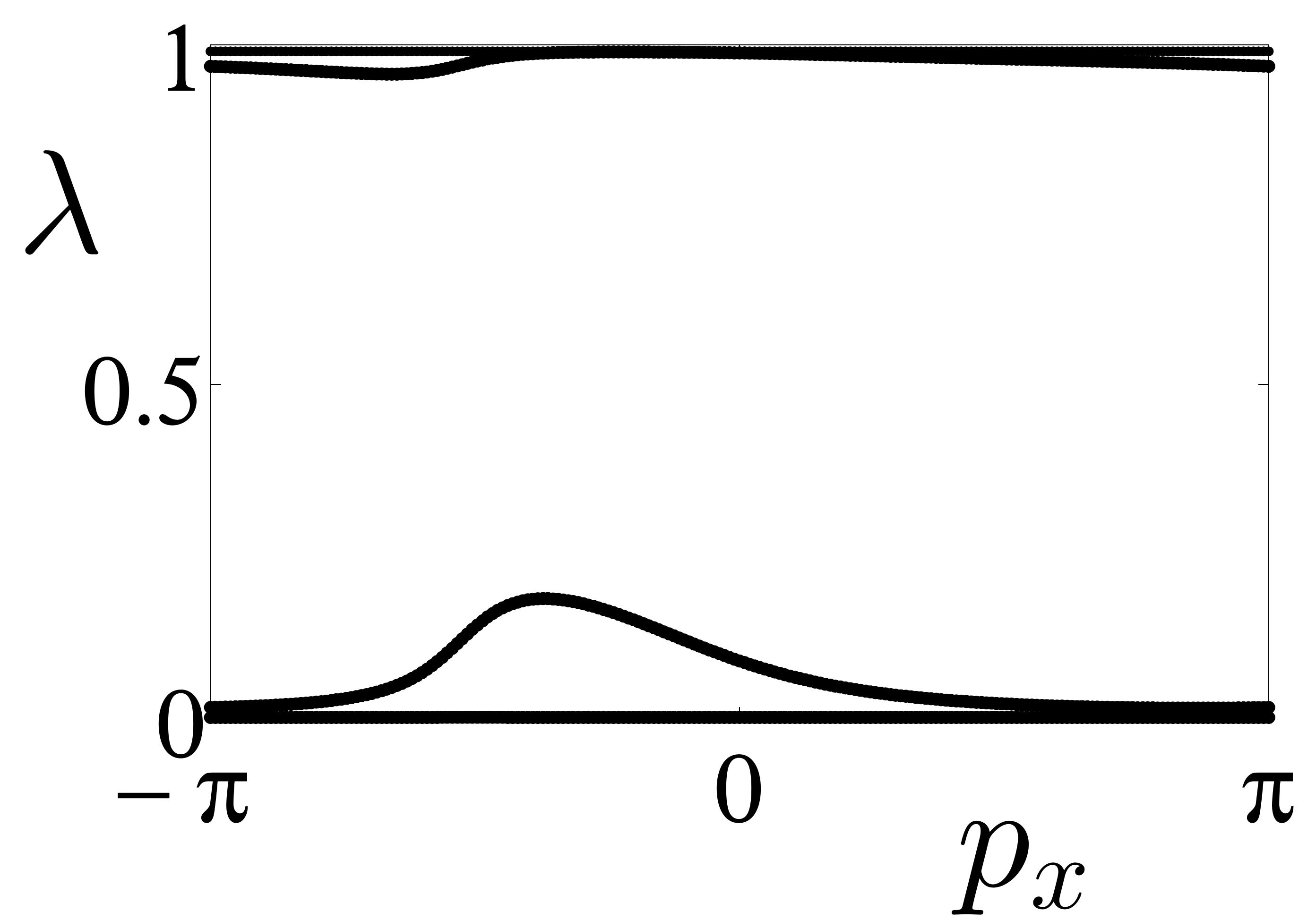}
\end{center}
\caption{\label{Fig4} (left) The gap-crossing line for $K=-J=1$, $m=0.5$ and $\theta=\pi/2$ shows the presence of an edge mode in the entanglement spectrum on the cylinder, that represents the signature of the physical edge state. (right) The absence of the crossing line for $K=-J=1$, $m\gg 1$ and $\theta=\pi/2$ is due to the absence of edge states in the topologically trivial insulating phase.}
\end{figure}
Let us briefly summarize the main properties of the entanglement spectrum for free-fermion systems. First of all, we divide our model which is placed on a cylinder in two non-overlapping sub-regions through a cut that runs along $x$, the periodic dimension \cite{Hermanns}. 
The reduced density matrix of the ground state $\hat{\rho}$ of the sub-system is related to the entanglement Hamiltonian $\hat{H}$, i.e. $\hat{\rho} = e^{-\hat{H}}$,
and the corresponding eigenvalues $\xi_{i}$ contain the main information about topological phases. However, in the case of free fermion Hamiltonians, it is possible to prove that $\xi_{i}$ are in one-to-one correspondence with the eigenvalues $\lambda_{i}$ of the correlation matrix \cite{Peschel}
\begin{eqnarray}
C_{nm}= \mathrm{tr} (\hat{\rho}\, h_{n}^{\dagger}h_{m})=\langle gs | h_{n}^{\dagger}h_{m}|
gs \rangle,
 \end{eqnarray}
where $| gs \rangle$ is the ground state, the labels $m, n$ are restricted to be within the sub-system, and in our case, $h_{n}$ represent the fermion operators $a_{n}$, $b_{n}$ and $c_{n}$. Thus, the entanglement spectrum in the model is the spectrum of the correlation matrix. Since our cut runs along the periodic direction $x$ we plot the eigenvalues $\lambda(p_x)$ of the correlation matrix $C(p_x)$ of each of the decoupled chains as showed in Fig. 4, where the ground state used to construct the correlation matrix is built by filling the lower band, and the entanglement cut is placed in the middle of each chain. In this figure, the virtual mode in the entanglement spectrum in the plot on the left represented by a gap-crossing line is the signature of the presence of a physical dispersing edge mode that will appear in the energy spectrum when a physical boundary is introduced in the place of the entanglement cut. This edge state is associated to the completely filled lower band which shows $\nu=1$. Instead, on the plot on the right, clearly the gap-crossing line disappears in the topologically trivial insulating phase when the band has $\nu=0$.\\

{\bf \em Disorder:--} In this section, we show that the Chern semimetals are even robust phases in presence of disorder, which would at least weakly appear in the case of experimentally realising the model with cold atoms in an optical lattice.
We impose periodic boundary conditions and devide the systems into four regions: three non-overlapping regions $X,Y,Z$ with common boundaries that form a tripple point which are all surrounded by the fourth region.
We then numerically calculate the Chern number in real space \cite{Kitaev, Bellissard} 
\begin{eqnarray}
\nu=12\pi i \sum_{j\in X}\sum_{k\in Y}\sum_{l\in Z} \left(   C_{jk}C_{kl}C_{lj} - C_{jl}C_{lk}C_{kj}  \right)
 \end{eqnarray}
which gives the same value as equation (\ref{Chern}) in the translationally invariant case. Numerically we are restricted to finite system sizes. However we use system size scaling in order to see that the real space Chern number converges to an integer as the system size is increased and thus choose an acceptable system size in order to perform our computation. Disorder is introduced in all of the hopping amplitudes $T=K,J,m$ in the form $T (1\pm J_d)$ where $J_d$ is the disorder amplitude. For each value of $J_d$ we compute the Chern number for a number of disorder realisations and average $\nu$ over them. The result as a function of $\theta$ and $J_d$ is shown in Fig. 5 (up).
\begin{figure}[htp]
\begin{center}
  \includegraphics[scale=0.195]{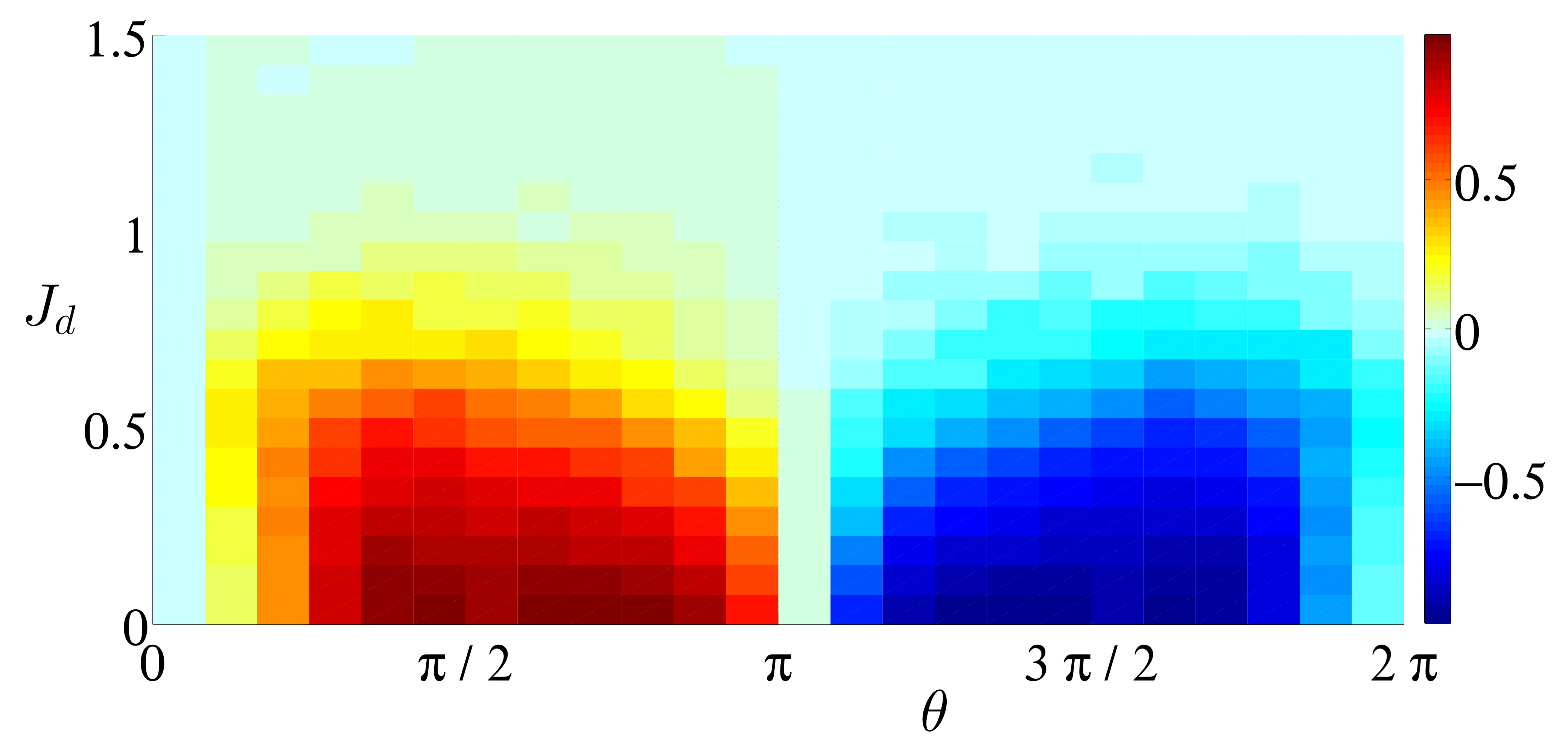}
\includegraphics[scale=0.165]{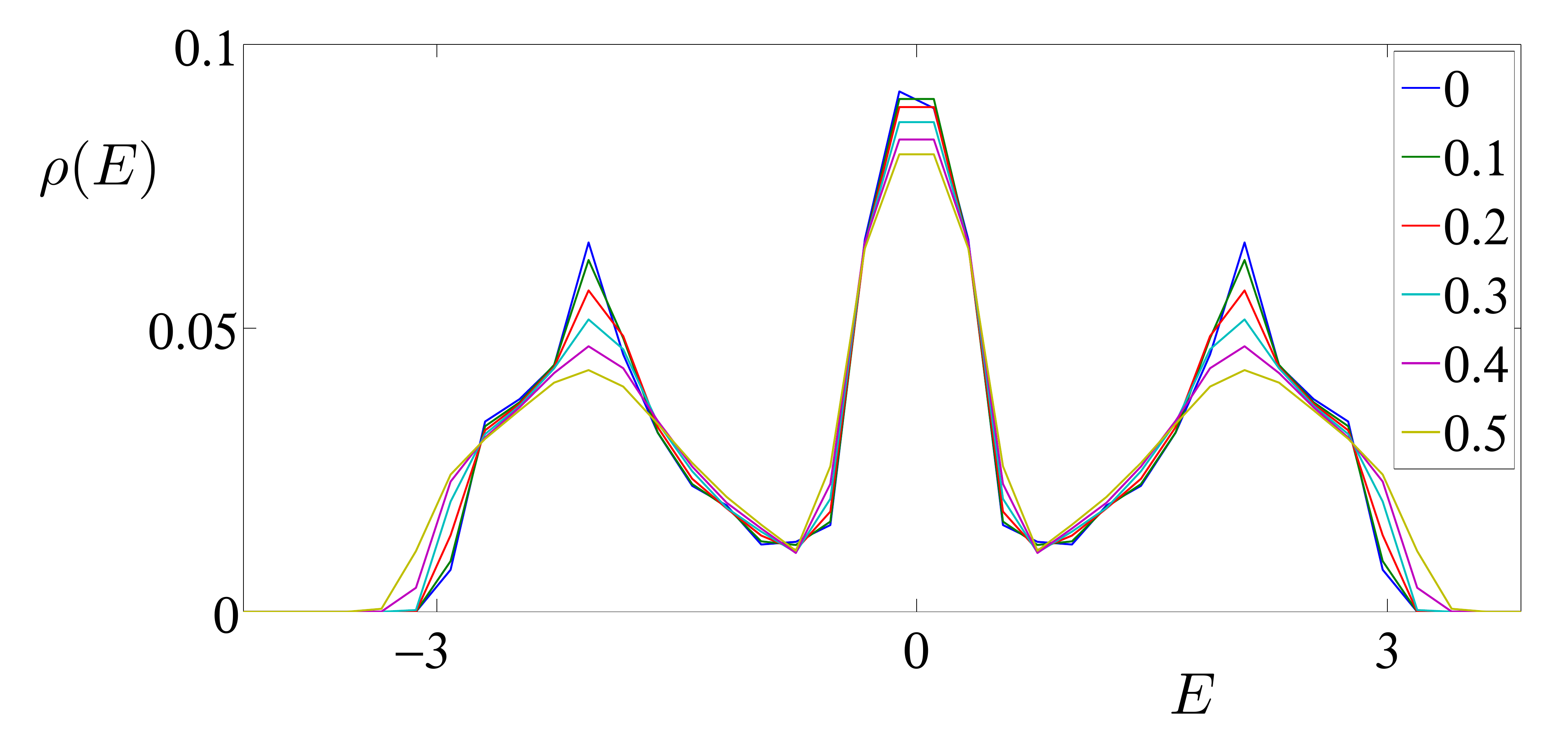}
\end{center}
\caption{\label{Fig5} (Up) Real space Chern number $\nu$ against $\theta$ and $J_d$ for $K=-J=1$ and $m=0.5$. Here a system size of $20 \times 20$ unit cells is chosen and the particle filling is what corresponds to filling the lowest band in the clean case. Each disorder realisation was averaged $20$ times, enough to discern the topological regions from the trivial ones. For weak disorder $J_d<0.5$ the topological phase is robust and for strong disorder the Chern number vanishes. (Low) Density of states $\rho(E)$ for disorder amplitudes $0<J_d<0.5$ (inset) where we averaged $10^3$ disorder realisations on $26\times 26$ unit cells for each $J_d$. The two minima correspond to the semimetalicl pseudo-gap. The density of states profile does not change around the pseudo-gaps. Here $K=-J=1$, $m=0.5$ and $\theta=\pi/2$.}
\end{figure}
For weak disorder, $J_d<1$ the topological phase is robust showing non-trivial value of the Chern number. Since there is no energy gap in the semimetallic phase, we look at the density of states  in order to identify the pseudo-gaps, where we have averaged a number of disorder realisations for each disorder amplitude Fig. 5 (low). In the clean case the density of states shows minima at the energies where the pseudo-gaps are located. The minima remain for weak disorder and we only notice a spread in the extreme energies away from the pseudo-gaps. This behaviour agrees with the Chern number remaining $| \nu | \approx 1$ for weak disorder showing the robustness of the Chern semimetallic phase.\\

{\bf \em Conclusions:--} Summarizing, we have shown that a new and simple tight-binding model on the Lieb lattice with only intra-unit-cell and nearest-neighbor hopping terms supports Chern semimetallic phases. These new topological semimetals are characterized by a non-zero Chern number in the bulk and topologically protected gapless edge states. We have proved the existence of the latter by studying the entanglement spectrum on the cylinder.
Moreover, we have shown that the Chern semimetal is robust also in presence of weak disorder by calculating the Chern number in the real space. Finally, for its simplicity, our Lieb lattice model can be experimentally realized with cold atoms in an optical lattice \cite{Goldman2, Shen}.\\

{\bf \em Acknowledgments:--} We thank Jiannis K. Pachos and Chris N. Self for discussions and comments. We are supported by EPSRC.

\end{document}